\documentclass[aps,prb,twocolumn,superscriptaddress,10pt]{revtex4}

\usepackage{graphicx}
\usepackage{amsmath}
\usepackage{amssymb}
\usepackage{amsfonts}
\usepackage{bm}
\usepackage{psfrag}
\usepackage{pstricks}

\usepackage{color}

\begin{document}

\title{Atomistic analysis of the impact of alloy and well-width fluctuations on the electronic and optical properties of InGaN/GaN quantum wells}

\author{Stefan Schulz}
\affiliation{Photonics Theory Group, Tyndall National Institute,
Dyke Parade, Cork, Ireland} 
\author{Miguel~A. Caro}
\affiliation{Photonics Theory Group, Tyndall National Institute,
Dyke Parade, Cork, Ireland} \affiliation{Department of Physics,
University College Cork, Cork, Ireland} \affiliation{Department of
Electrical Engineering and Automation, Aalto University, Espoo,
Finland} \affiliation{COMP Centre of Excellence in Computational
Nanoscience, Aalto University, Espoo, Finland}
\author{Conor Coughlan}
\affiliation{Photonics Theory Group, Tyndall National Institute,
Dyke Parade, Cork, Ireland} \affiliation{Department of Physics,
University College Cork, Cork, Ireland}
\author{Eoin~P. O'Reilly}
\affiliation{Photonics Theory Group, Tyndall National Institute,
Dyke Parade, Cork, Ireland} \affiliation{Department of Physics,
University College Cork, Cork, Ireland}

\begin{abstract}
We present an atomistic description of the electronic and optical
properties of $\text{In}_{0.25}\text{Ga}_{0.75}$N/GaN quantum wells.
Our analysis accounts for fluctuations of well width, local alloy
composition, strain and built-in field fluctuations as well as
Coulomb effects. We find a strong hole and much weaker electron wave
function localization in InGaN random alloy quantum wells. The
presented calculations show that while the electron states are
mainly localized by well-width fluctuations, the holes states are
already localized by random alloy fluctuations. These localization
effects affect significantly the quantum well optical properties,
leading to strong inhomogeneous broadening of the lowest interband
transition energy. Our results are compared with experimental
literature data.
\end{abstract}

\date{\today}



\maketitle

Over the last twenty years, research into nitride-based
semiconductor materials (InN, GaN, AlN and their respective alloys)
has gathered pace. This stems from their potential to emit light
over a wide spectral range, making them highly attractive for
different applications.~\cite{Hump2008} Despite very high defect
densities, blue emitting $\text{In}\text{Ga}$N-based devices exhibit
high quantum efficiencies.~\cite{Naka98,OlBe2010} The widely
accepted explanation for this is that the carriers are spatially
localized due to alloy fluctuations and are thus prevented from
diffusing to defects.~\cite{Naka98,ChUe2006,ScHa2004,OlBe2010} The
impact of alloy fluctuations on the electronic and optical
properties in $c$-plane InGaN/GaN quantum wells (QWs) has been
further evidenced experimentally, e.g., by the ``S-shape''
temperature dependence of the peak photoluminescence (PL)
energy.~\cite{WaJi2012,HaWa2012} It is important to note that in
\emph{wurtzite} (WZ) InGaN systems the effect of these fluctuations
is much more severe compared to that found, e.g., in
\emph{zinc-blende} (ZB) InGaAs alloys. This originates from the very
different physical properties (e.g., band gap and lattice spacing)
of the binary constituents (InN and GaN).~\cite{Hump2008,OlBe2010} A
further complication is that InGaN/GaN QWs, compared with
InGaAs/GaAs wells, exhibit much stronger electrostatic built-in
fields, arising in part from the strain dependent piezoelectric
response.~\cite{BeFi97} Thus, alloy fluctuations in InGaN/GaN QWs
affect the electronic structure through a complicated interplay of
local alloy, strain, and built-in field fluctuations.

Even though the importance of alloy fluctuations has been
experimentally evidenced, they have been widely neglected in the
modeling of \emph{WZ} InGaN/GaN QWs. Previous \emph{atomistic}
calculations have mainly focused on
ZB~\cite{LeWa2006,WuWa2009,ChLi2010} or
WZ~\cite{GoLe2009,LiLu2010,MoMi2011,PeCa2011} InGaN \emph{bulk}
alloys. Properties of WZ InGaN/GaN QWs with realistic dimensions are
typically studied using continuum-based theoretical models, which
inherently overlook alloy or built-in field fluctuations on a
microscopic level. However, there are also some continuum-based
approaches to mimic the impact of alloy fluctuations on the
electronic and optical properties of WZ InGaN/GaN QWs. For example,
Funato and Kawakami~\cite{FuKa2008} modeled alloy fluctuations in a
continuum-based approach by taking lateral confinement effects into
account in the model, leading therefore to quantum dot (QD) like
structures. This introduces, however, the effect that electron and
hole wave functions are spatially localized at the same
\emph{in-plane} position. In a microscopic description of a random
alloy, this need not necessarily be the case as we will show here.
Watson-Parris and co-workers~\cite{WaGo2011} analyzed, based on a
single-band effective mass approximation (EMA), the impact of alloy
fluctuations on the electronic and optical properties of $c$-plane
InGaN/GaN QWs by assuming that the material parameters vary
spatially. A similar approach has been recently applied by Yang
\emph{et al.}~\cite{YaSh2014} The authors highlighted the importance
of alloy fluctuations for an accurate modeling of these
systems.~\cite{WaGo2011,YaSh2014} However, these continuum-based
models overlook the underlying atomistic (anion-cation) structure, a
feature that has been shown to be important for, e.g., AlInN
alloys.~\cite{ScCa2013Apex} Also, the chosen single-band EMA of
Ref.~\onlinecite{WaGo2011} does not account for valence-band (VB)
mixing effects. Moreover, the stronger carrier localization expected
in regions containing In chains and clusters,~\cite{LiLu2010} is
overlooked in a continuum-based description.

Here, we provide microscopic insight into the impact of alloy and
well width fluctuations on the electronic and optical properties of
InGaN/GaN QWs. We take as an example a series of
$\text{In}_{x}\text{Ga}_{1-x}$N/GaN QWs with 25 \% InN content
($x=0.25$). Our electronic structure model is based on an atomistic
tight-binding (TB) model, taking input (local strain and
electrostatic built-in fields) from our recently established local
polarization theory.~\cite{CaSc2013local} This framework has already
been validated against both density functional theory (DFT) and
experimental data,~\cite{CaSc2013local,ScCa2013Apex} showing an
excellent agreement between our semi-empirical theory, DFT, and
experiment. Coulomb effects are treated in the configuration
interaction (CI) scheme based on the calculated electron and hole TB
wave functions, thus taking mixing between different states into
account.

We show here, based on our microscopic approach, that the assumption
of a random InGaN alloy in the QW region leads already to strong
hole wave function localization effects. These effects are less
pronounced for electron states. However, as we demonstrate, the
ground-state electron wave function is strongly affected by the
presence of well-width fluctuations (WWFs). Our analysis reveals
also that \emph{local} strain effects, arising from local alloy
fluctuations, lead to a situation where different microscopic alloy
configurations lead to very different orbital mixing effects into
the hole ground state.

Additionally, we discuss not only ground-state properties but also
excited states in InGaN/GaN QWs. These excited states are important
for explaining the experimentally observed ``S-shape'' temperature
dependence of the PL peak energy. We demonstrate by explicit
calculation that also excited hole states are strongly localized.

Finally, we compare our full model, including Coulomb effects, with
available experimental data. Our theoretical results and trends are
in good agreement with values reported in the literature.

The paper is organized as follows. In the following section, we
introduce the ingredients of our theoretical framework.
Section~\ref{sec:QWModel} describes, based on available experimental
literature data, the QW structure under consideration. Our results
are presented in Sec.~\ref{sec:Results}. In
Sec.~\ref{sec:Results_Experiment}, we compare the obtained results
with experimental data from the literature. Finally, we summarize
our work in Sec.~\ref{sec:Summary}.

\section{Theoretical Framework}
\label{sec:theory}

In this section, we introduce the microscopic theoretical framework
we use to study the electronic and optical properties of InGaN/GaN
QWs. In a first step, Sec.~\ref{sec:StrainPotential}, we describe
our atomistic strain field and built-in potential model. In
Sec.~\ref{sec:TBModel}, the electronic structure theory, based on a
TB model, is introduced. Finally, Sec.~\ref{sec:ManyBdy} deals with
the calculation of the optical properties of InGaN/GaN QWs by means
of a CI scheme.

\subsection{Strain field calculations, local polarization theory, and local built-in potential model}
\label{sec:StrainPotential}

Macroscopic electric polarization in an insulating heteropolar
material arises from a non-vanishing sum of electric dipoles which
is, in turn, a consequence of the lack of inversion symmetry in the
material. This lack of inversion symmetry and the resulting electric
polarization can already be present in the unstrained sample (e.g.,
spontaneous polarization in the WZ lattice) but it can also be
enhanced by applied strain. For binary compounds, such as pure GaN,
macroscopic strain induces an equivalent microscopic deformation of
the unit cell, and the link between local and macroscopic electric
polarization can be established.~\cite{CaSc2013local} In the case of
alloyed compounds, however, macroscopic strain cannot be directly
related to the local deformation of the crystal. Take as an example
group-III nitrides: since the In--N bond distances in InN are larger
(by about $\sim$10\%) than the Ga--N bond distances in GaN, each of
the atomic tetrahedra with an In atom in the middle in an InGaN
alloy will be compressively strained while those with a Ga atom in
the middle will undergo tensile stretching. The exact extent of this
\textit{local} strain varies throughout the crystal depending on the
specific local atomic configuration. In order to take the effects of
local strain and disorder on electric polarization into account a
local theory of polarization is needed. We have already presented
the foundations of this theory and an assessment of its degree of
applicability for group-III nitrides.~\cite{CaSc2013local} The
theory relies on decoupling the macroscopic and local contributions
to the electric polarization, where the latter can be evaluated at
each atomic site. The corresponding expression for the $i\text{th}$
component of the \textit{local} polarization vector field is given
by
\begin{align}
P_i = & \underbrace{\sum\limits_{j=1}^6 e_{ij}^{(0)}
\epsilon_j}\limits_\text{macroscopic}
\nonumber \\
& + \underbrace{ P_i^\text{sp} - \frac{e}{V_0}
\frac{\mathcal{Z}^0_i}{N_\text{coor}^0} \left( \mu_i -
\sum\limits_{j=1}^3 \left( \delta_{ij} + \epsilon_{ij} \right)
\mu_{j,0} \right) }\limits_\text{local}, \label{mig-01}
\end{align}
where $e_{ij}^{(0)}$ are the clamped-ion piezoelectric coefficients,
$\epsilon_j$ (in Voigt notation; $\epsilon_{ij}$ in cartesian
notation) are the macroscopic strain components, $P_i^\text{sp}$ is
the spontaneous polarization and $V_0$ is the volume assigned to
each atomic site (for tetrahedrally-bonded crystals, 6 times the
volume of a tetrahedron). The elementary charge is denoted by $e$,
$\mathcal{Z}^0_i$ is the Born effective charge of the atom at whose
site the local polarization is being computed, and $N_\text{coor}^0$
is its number of nearest neighbors. The parameter $\mu_i$ arises
from a sum over nearest-neighbor distances ($\mu_{j,0}$ is this same
parameter before strain) and $\delta_{ij}$ is the Kronecker delta.
The derivation of Eq.~(\ref{mig-01}) and further detail on the
meaning and significance of all the quantities involved can be found
in Ref.~\onlinecite{CaSc2013local}.

The accuracy that can be attained with Eq.~(\ref{mig-01}) relies
greatly on the level of theory employed to obtain the different
parameters that appear in the expression. Except for $\epsilon_j$
and $\mu_i$, all of them can be calculated independently of the size
of the system and be transferred across calculations. The parameters
for the III-N compounds have been calculated on the basis of density
functional theory (DFT) within the Heyd-Scuseria-Ernzerhof (HSE)
screened exchange hybrid functional scheme and can be found in
Ref.~\onlinecite{CaSc2013local}. Macroscopic strain $\epsilon_j$ and
the asymmetry parameter $\mu_i$ are system specific and require an
explicit evaluation for the system at hand. For large systems, such
as random alloy InGaN/GaN QWs with WWFs of realistic size as studied
here, calculations at the DFT level are unaffordable and an
alternative \textit{atomistic} approach is required. For homopolar
tetrahedrally-bonded compounds, commonly available force fields are
based on the valence force field (VFF) derived by Musgrave and Pople
for diamond,~\cite{MuPo1962} or the Keating
potential.~\cite{Keating66} For heteropolar tetrahedrally-bonded
compounds, notably ZB, Martin proposed a generalization of both
models to include electrostatic interactions
explicitly.~\cite{Martin70} Martin's expression for the total energy
of atom $i$ in the ZB unit cell, including VFF and electrostatic
contributions, is given by
\begin{align}
U_i = & \frac{1}{2}\sum\limits_{j \neq i} \frac{1}{2} k_r (r_{ij} -
r_{ij}^0)^2
\nonumber \\
& + \sum\limits_{j \neq i} \sum\limits_{k \neq i, k>j} \Bigl\{
\frac{1}{2} k_\theta^i r_{ij}^0 r_{ik}^0 (\theta_{ijk} -
\theta_{ijk}^0)^2
\nonumber \\
& + k_{r \theta}^i \left[ r_{ij}^0 (r_{ij} - r_{ij}^0) + r_{ik}^0
(r_{ik} - r_{ik}^0) \right] (\theta_{ijk} - \theta_{ijk}^0)
\nonumber \\
& +  k_{r r}^i (r_{ij} - r_{ij}^0) (r_{ik} - r_{ik}^0) \Bigr\}
\nonumber \\
& + {\sum\limits_{j \neq i}}' \frac{Z_i^* Z_j^* e^2}{4 \pi
\epsilon_r \epsilon_0 r_{ij}} - \frac{1}{2} \sum\limits_{j \neq i}
\frac{1}{4} \alpha_M \frac{Z_i^* Z_j^* e^2}{4 \pi \epsilon_r
\epsilon_0 {r_{ij}^0}^2} (r_{ij} - r_{ij}^0)\, . \label{mig-02}
\end{align}
The different $k_r$, $k_\theta^i$, $k_{r \theta}^i$, and $k_{r r}^i$
denote the force constants. The angle between atoms $i$, $j$ and $k$
is given by $\theta_{ijk}$, $Z_i^*$ denotes the effective charge of
atom $i$ in a point charge model (that can be positive or negative)
and $e$ is the elementary charge. The permittivity of the vacuum is
given by $\epsilon_0$, while $\epsilon_r$ denotes the dielectric
constant of the material. The Madelung constant is $\alpha_M$, which
in the case of the ZB lattice is given by $\alpha_M = 1.6381$. The
last term in Eq.~(\ref{mig-02}) is a linear repulsion term, required
for the crystal to be stable,~\cite{Martin70} that counteracts the
linear elements obtained from the power expansion of the
electrostatic part of the energy. All summations run over the
first-nearest neighbors of atom $i$, except the summation marked
with a prime symbol, corresponding to the long-ranged Coulomb
interaction, which runs over the whole crystal. To avoid double
counting over atoms in the same unit cell, a factor $\frac{1}{2}$
has been introduced for the two summations involving bond-stretching
terms. Martin has also given the relation between the force
constants of the general VFF in Eq.~(\ref{mig-02}) and Keating's
potential. The main advantage of using Eq.~(\ref{mig-02}) for WZ is
that the inclusion of the electrostatic terms leads to the important
qualitative result of a $c/a$ ratio and internal parameter $u$ that
deviate from the ideal values ($\sqrt{8/3}$ and $3/8$,
respectively). We have implemented Eq.~(\ref{mig-02}) in the
software package \texttt{gulp}.~\cite{GaRo2003} Fitting of the
different force constants and the effective charges to structural
and elastic properties of the WZ material in question leads also to
good quantitative description of those quantities. More details of
our VFF model will be given elsewhere.

Having established the local polarization vector field,
Eq.~(\ref{mig-01}), and the underlying VFF, Eq.~(\ref{mig-02}), in a
final step one needs to calculate the corresponding \emph{local}
built-in polarization potential $\phi$. As discussed in detail in
Ref.~\onlinecite{CaSc2013local}, we perform these calculations on
the basis of a point dipole method. The point dipole model is a
solution to the challenge of solving Poisson's equation on an
atomistic grid, where abrupt changes in the polarization occur.

\subsection{Electronic structure calculations}
\label{sec:TBModel}

To model the impact of local alloy fluctuations on the electronic
and later on the optical properties of InGaN/GaN QWs by means of an
atomistic approach, we choose here an $sp^3$ TB model. The TB
parameters at each atom site $\mathbf{R}$ of the underlying WZ
lattice are set according to the bulk values of the respective
occupying atom. Here, the bulk TB parameters are obtained by fitting
the TB band structures of InN and GaN to the corresponding HSE
hybrid-functional DFT results, as described in detail in
Ref.~\onlinecite{CaSc2013local}.

Since for the cation sites (Ga, In) the nearest neighbors are always
nitrogen atoms, there is no ambiguity in assigning the TB on-site
and nearest neighbor matrix elements. This classification is more
difficult for the nitrogen atoms. In this case the nearest neighbor
environment is a combination of In and Ga atoms. Here, we apply the
widely used approach of using weighted averages for the on-site
energies according to the number of In and Ga
atoms.~\cite{OReLi2002,LiPo92,BoKh2007}

In setting up the Hamiltonian, one has to include the local strain
tensor $\epsilon_{ij}(\mathbf{r})$ and the local built-in potential
$\phi(\mathbf{r})$ to ensure an accurate description of the
electronic properties of the InGaN alloy. Several authors have shown
that strain effects can be introduced by on-site corrections to the
TB matrix elements
$H_{l\mathbf{R}',m\mathbf{R}}$,~\cite{JaSc98,BoKl2002} where
$\mathbf{R}$ and $\mathbf{R}'$ denote lattice sites and $l$ and $m$
are the orbital types. Here, we include the strain dependence of the
TB matrix elements via the Pikus-Bir
Hamiltonian~\cite{WiSc2006,ScBa2010} as a site-diagonal
correction.~\cite{CaSc2013local} With this approach, the relevant
deformation potentials for the highest valence and lowest conduction
band states at the $\Gamma$ point are included directly without any
fitting procedure. The deformation potentials for InN and GaN are
taken from HSE-DFT calculations.~\cite{YaRi2009} Again on the same
footing as in the case of the on-site energies for the nitrogen
atoms we use weighted averages to obtain the strain dependent
on-site corrections for $\text{In}_{x}\text{Ga}_{1-x}$N. Our
approach is similar to that used for the strain dependence in an
8-band $\mathbf{k}\cdot\mathbf{p}$ model,~\cite{WiSc2006} but has
the benefit that the TB Hamiltonian is sensitive to the distribution
of local In, Ga and N-atoms.

The last ingredient to our TB model for the description of the
electronic structure of $\text{In}_{x}\text{Ga}_{1-x}$N systems is
the local built-in potential $\phi(\mathbf{r})$ arising from
piezoelectric and spontaneous polarization contributions as
discussed in Sec.~\ref{sec:StrainPotential}. The built-in potential
$\phi(\mathbf{r})$ is likewise included as a site-diagonal
contribution in the TB
Hamiltonian.~\cite{RaAl2003,SaAr2002,ZiJa2005,ScBa2012}

\subsection{Many-body calculations}
\label{sec:ManyBdy}

Having discussed the TB Hamiltonian used for the description of the
QW single-particle states, we now turn our attention to the
investigation of the optical properties of the studied QW system. In
a first step, we discuss the calculation of the interaction matrix
elements. In a second step, we outline the CI scheme and the
calculation of the optical spectra.

\subsubsection{Calculation of interaction matrix elements}

For the calculation of optical spectra, Coulomb and dipole matrix
elements between TB single-particle wave functions are required. As
the atomic orbitals are not explicitly known in an empirical TB
approach, we approximate the Coulomb matrix elements
by:~\cite{ShCh2005,ScSc2006,ZiMa2014}
\begin{align}\label{EqCoulappr}
&V_{ijkl}=
\sum_{\mathbf{R}\mathbf{R}'}\sum_{\alpha\beta}c_{\mathbf{R}\alpha}^{i\ast}
c_{\mathbf{R}'\beta}^{j\ast}c^k_{\mathbf{R}'\beta}c^l_{\mathbf{R}\alpha}V(\mathbf{R}-\mathbf{R}')\,,\\\nonumber
&\text{with}\quad V(\mathbf{R}-\mathbf{R}')=
\frac{e^2}{4\pi\epsilon_0\epsilon_r|\mathbf{R}-\mathbf{R}'|}\quad\text{for}
\quad \mathbf{R}\not=\mathbf{R}'\\\label{EqCoulappr1} &\text{and}
\quad
V(0)=\frac{1}{V^2_{uc}}\int_{uc}\mathrm{d}^3\mathbf{r}\,\mathrm{d}^3\mathbf{r}'\frac{e^2}{4\pi\epsilon_0\epsilon_r|\mathbf{r}-\mathbf{r}'|}\approx
\tilde{V}_0 \,.
\end{align}
The $c^i_{\mathbf{R}\alpha}$ are the expansion coefficients of the
${i^{th}}$ TB single-particle wave function
$\psi_i(\mathbf{r})=\sum_{\mathbf{R}\alpha}c_{\mathbf{R}\alpha}^i\Phi_{\mathbf{R}\alpha}(\mathbf{r})$,
in terms of the atomic orbitals
$\Phi_{\mathbf{R}\alpha}(\mathbf{r})$ localized at the position
$\mathbf{R}$. In Eq.~(\ref{EqCoulappr}) the variation of the Coulomb
interaction is taken into account only on a length scale of the
order of the lattice vectors but not inside one unit cell. This is
well justified due to the long ranged, slowly varying behavior of
the Coulomb interaction. For $|\mathbf{R}-\mathbf{R}'|=0$ the
evaluation of the integral in Eq.~(\ref{EqCoulappr1}) can be done
quasi-analytically by expansion of the Coulomb interaction in terms
of spherical harmonics.~\cite{ScCz2002} The details can be found in
Ref.~\onlinecite{ScSc2006}.

However, in an alloyed system this approach becomes more difficult
for the on-site matrix elements as the size of the unit cell changes
depending on the local environment. To simplify this approach we
work here with $\approx$16 eV for the \emph{unscreened} on-site
Coulomb matrix elements.~\cite{ScSc2006} This value is in accordance
with other calculations for this type of matrix
elements.~\cite{ScCz2002} However, when taking screening effects
into account and assuming a linearly interpolated dielectric
constant for InGaN alloys, as assumed in Ref.~\onlinecite{FuKa2008},
$\tilde{V}_0$ is of the order 1-2 eV. To test the impact of the
unscreened on-site matrix elements on the results we have changed
its value by 4 eV to 12 eV. We find here that the direct Coulomb
matrix element $V^{eh}_{1111}$ is affected by less than 0.5 meV when
changing the on-site Coulomb matrix element by 4 eV. The origin of
this is related to the fact that the direct Coulomb interaction is
dominated by long range contributions as discussed above. Therefore,
changes and local variations of the on-site Coulomb matrix elements
should be of secondary importance.

Furthermore, the Coulomb interaction is evaluated on one supercell
only. As we will see later, the wave functions of electrons and
holes are strongly localized inside the supercell.

In contrast to the Coulomb matrix elements, the short range
contributions dominate the dipole matrix elements. Thus, it is
necessary to connect the calculated TB coefficients directly to the
underlying set of atomic orbitals. A commonly used approach is the
use of Slater orbitals.~\cite{LeJoe2001} These orbitals include the
correct symmetry properties of the underlying TB coefficients but
lack the essential assumption of orthogonality with respect to
different lattice sites, since they have been developed for isolated
atoms. We have previously overcome this problem by using numerically
orthogonalized Slater orbitals.~\cite{ScSc2006} These orthogonalized
orbitals fulfill all basic requirements, regarding the symmetry,
locality, and orthogonality of the basis orbitals underlying the TB
formulation. However, in general we can decompose the dipole
operator into an envelope part and an orbital part. Here, we perform
the calculation of the envelope part only, since we are only
interested in getting first insights into the relative strength of
different transitions from different microscopic configurations.

\subsubsection{Configuration interaction scheme and optical spectra}

In this section we briefly describe the CI approach and the
calculation of the optical spectra. More details on the CI scheme
are given, for example, in
Refs.~\onlinecite{BaDu95,FrFu99,BaGa2004,ScSc2006}. Since we are
dealing with strongly localized states, as we will see later, we use
our approaches developed for QD systems.~\cite{ScSc2006}

In the CI calculation the microscopically evaluated single-particle
states and Coulomb interaction matrix elements serve as an input to
determine the many-body eigenstates. To this end, the Hamiltonian is
expressed in terms of all possible Slater determinants that can be
constructed for the finite localized single-particle basis for a
given number of electrons and holes. In the following we are
interested in effects arising from one electron-hole pair.
Therefore, electron-electron and hole-hole Coulomb interactions are
not required. We neglect here electron-hole exchange contributions
since these are small corrections on the energy scale relevant for
the discussion of our results. The resulting many-body matrix is
diagonalized and Fermi's golden rule is used to evaluate dipole
transitions between the Coulomb-correlated
states:~\cite{BaDu95,FrFu99,BaGa2004}

\begin{equation}
\label{eq:Fermi} I(\omega)={2\pi \over \hbar} \sum_f |\langle
\phi_f| H_{\text{D}} |\phi_i \rangle |^2 \,\,
\delta(E_i-E_f-\hbar\omega) \quad .
\end{equation}
Here $|\phi_i\rangle$  denotes the correlated initial state with
energy $E_i$ and $|\phi_f\rangle$ and  $E_f$ the corresponding
quantities of the final states. A similar equation holds for the
absorption spectrum. The Hamiltonian $H_{\text{D}}$ describes the
light matter interaction in dipole approximation
\begin{equation}
\label{eq:Hdipol} H_{\text{D}} = -\sum_{n,m} \mathbf{E}
\mathbf{d}^{eh}_{nm} h^{\dag}_{n} e^{\dag}_{m}  + \text{h.c.}
\,\quad{,}
\end{equation}
where $h^{\dag}_{n}$ and $e^{\dag}_{m}$ are hole and electron
creation operators, respectively. In this expression
$\mathbf{d}^{eh}_{nm}$ denotes the dipole matrix elements $\langle
n|e\mathbf{r}|m\rangle$ with the single-particle states $|n\rangle$
and $|m\rangle$ for the electron and hole, respectively. The
quantity $\mathbf{E}$ is the electric field at the position of the
QW and $e$ is the elementary charge. Here, we assume that the
polarization vector $\mathbf{e}_p$ of the electric field is given by
$\mathbf{e}_{p}=\frac{1}{\sqrt{2}}(1,1,0)^{t}$, which corresponds to
standard experimental set up.~\cite{HaWa2012,WaJi2012} Fermi's
golden rule, Eq.~(\ref{eq:Fermi}), shows that the optical field
always creates or destroys electron-hole pairs. From this it is
immediately obvious that the only non-zero transition will stem from
situations where the initial and final state differ by exactly one
electron-hole pair.

\begin{figure}[t]
\includegraphics{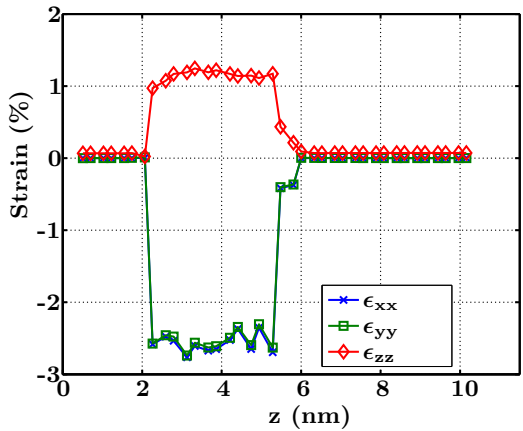}
\caption{(Color online) Average strain tensor components
$\epsilon_{xx}$, $\epsilon_{yy}$ and $\epsilon_{zz}$ along the z
direction (c axis).} \label{fig:Strain}
\end{figure}

\section{Model system}
\label{sec:QWModel}

Having described the ingredients of our theory, we now introduce the
QW structure being considered. As a model system we assume an
approximately 3.5 nm wide $\text{In}_{0.25}\text{Ga}_{0.75}$N/GaN
QW. This structure is similar to the experimentally studied system
in Ref.~\onlinecite{GrSo2005}. All QW calculations have been
performed on supercells containing $\approx$82,000 atoms ($\approx10
\text{nm} \times 9 \text{nm} \times 10 $nm) with periodic boundary
conditions. Following the experimental data in
Refs.~\onlinecite{MoPa2002,SmKa2003,GaOl2007,GaOl2008,BeSa2011,BaJo2013},
we treat InGaN as a random alloy. To realize different microscopic
configurations, our calculations have been repeated ten times with
changing the atomic distribution. Furthermore, experimental studies
reveal WWFs at the upper QW interface.~\cite{GaOl2008,MoVi2009} The
diameter of these well width fluctuations is $\approx$5-10 nm, while
their height is between one and two monolayers. To treat such
fluctuations, we assume disklike WWFs with a diameter of 5 nm and a
height of two MLs, residing on the
$\text{In}_{0.25}\text{Ga}_{0.75}$N/GaN QW.

\section{Results}
\label{sec:Results}

In this section we analyze strain fields, built-in potentials, and
the electronic and optical properties of the QW structure under
consideration. In a first step we discuss the strain field and the
built-in potential in the QW structure, including local effects and
WWFs. In Sec.~\ref{sec:Results_Electronic} we address the electronic
structure of the QW, while Sec.~\ref{sec:Results_Optical} deals with
the impact of Coulomb effects on the results.

\subsection{Strain field and built-in potentials}

In Fig.~\ref{fig:Strain} we show, following the approach by Pryor
\emph{et al.}~\cite{PrKi98}, the strain tensor components
$\epsilon_{xx}$, $\epsilon_{yy}$ and $\epsilon_{zz}$ along the $c$
axis ($z$ axis) for one of the structures considered. The data shown
here are averaged over the whole supercell. Several features are
clearly visible. To a first approximation the different components
reflect the profile one would expect from a continuum-based
description.~\cite{CaSc2011} However, even when averaging the strain
tensor components over the supercell, the impact of local alloy
fluctuations giving rise to local strain tensor fluctuations is
clearly visible. As we have seen already for random InGaN
\emph{bulk} systems, local strain field fluctuations significantly
affect the valence and conduction band edges.~\cite{CaSc2013local}
Also the features arising from the WWFs are visible in the region of
$z\approx5.5-6$ nm.

\begin{figure}[t]
\includegraphics{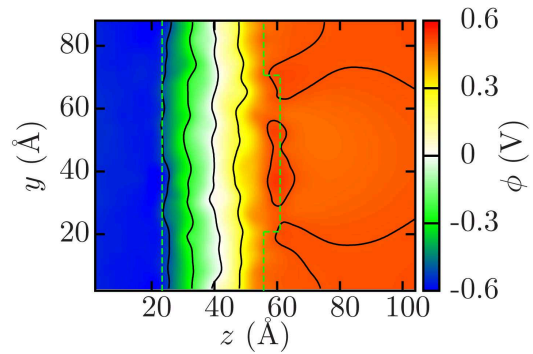}
\caption{(Color online) Contour plot of the built-in potential
$\phi$ of an $\text{In}_{0.25}\text{Ga}_{0.75}$N/GaN QW for a slice
through the center of the cylindrical shaped WWF in the $y-z$-plane.
The dashed lines indicate the QW interfaces. The $z$ direction is
parallel to the $c$ axis.} \label{fig:potential_plot}
\end{figure}

In a second step, we discuss the result from our local polarization
theory. Figure~\ref{fig:potential_plot} displays the calculated
built-in potential $\phi$ for a slice through the center of the
cylindrical-shaped WWF in the $y-z$-plane for one of the considered
structures. Again, the $z$ direction is parallel to the $c$ axis. As
we can see from Fig.~\ref{fig:potential_plot}, the isolines inside
the QW are \emph{not} straight lines. This is in contrast to a
continuum-based description. In our atomistic approach, the isolines
are affected by the local strain and alloy fluctuations. In
addition, the shape of the built-in potential is significantly
affected by the presence of the WWF.

\begin{figure}[t]
\includegraphics{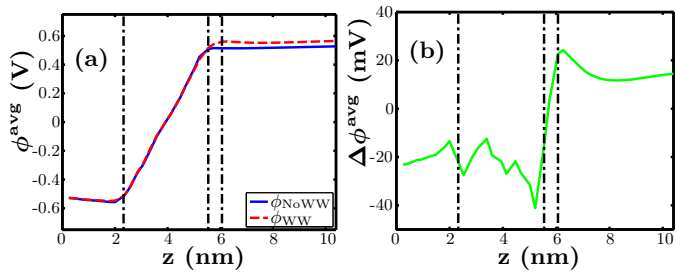}
\caption{(Color online) (a) Averaged built-in potential for a
line-scan through the QW along the $c$-axis. Results with
($\phi_\text{WW}$) and without ($\phi_\text{NoWW}$) WWFs are shown.
Dashed-dotted lines indicate approximately the QW interfaces. (b)
$\Delta \phi^\text{avg}=\phi_\text{WW}-\phi_\text{NoWW}$.}
\label{fig:potential}
\end{figure}

To further analyze the impact of WWFs on the built-in potential
$\phi(\mathbf{r})$ in $c$-plane InGaN/GaN QWs,
Fig.~\ref{fig:potential} (a) shows the built-in potential
$\phi^\text{avg}$, averaged over the ten different configurations,
for a line-scan through the QW along the $c$ axis. The line-scan
runs through the center of the disklike WWF. $\phi^\text{avg}$
reflects to a first approximation the capacitorlike behavior one
would expect from a continuum-based description. The solid line
shows the result in the absence of WWFs ($\phi_\text{NoWW}$), while
the dashed line displays the built-in potential in the presence of
WWFs ($\phi_\text{WW}$). The vertical dashed-dotted lines indicate
the QW interfaces along the $c$ axis.

To analyze the impact of alloy and WWFs on the built-in potential in
more detail, Fig.~\ref{fig:potential} (b) depicts $\Delta
\phi^\text{avg}=\phi_\text{WW}-\phi_\text{NoWW}$. From this we can
conclude two things. Firstly, even when averaging over the ten
configurations the influence of the alloy fluctuations on built-in
potential is clearly visible since $\Delta \phi^\text{avg}$ is
clearly not smooth in the QW region. Secondly, WWFs lead to a
reduced built-in field inside the InGaN/GaN QW and near the upper
interface. The change in the slope of $\phi^\text{avg}$ inside the
QW leads to the effect that the electron wave function can leak
further into the QW center. The origin of the built-in field
reduction can be explained using linear continuum elasticity theory.
In this approach, the total built-in potential is the sum of the
potential arising from the QW plus a contribution arising from a
disk-shaped QD. By looking at Fig.~\ref{fig:potential} (b), the
profile of $\Delta \phi^\text{avg}$ reflects the characteristic of
nitride-based QDs.~\cite{ScORe2011}

\begin{figure}[t!]
\includegraphics{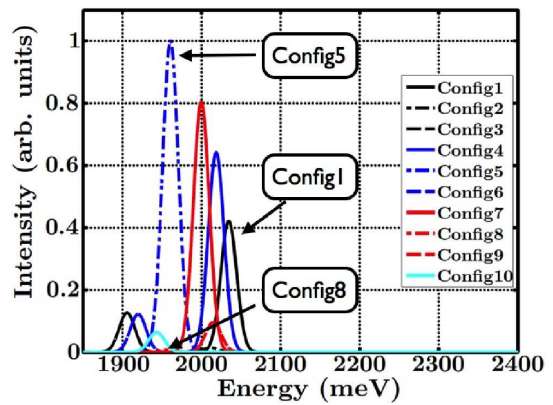}
\caption{(Color online) \emph{Single-particle} ground-state emission
spectrum of an $\text{In}_{0.25}\text{Ga}_{0.75}$N/GaN QW for
different configurations including random alloy and well width
fluctuations.} \label{fig:Spectrum_WW_SP}
\end{figure}

\subsection{Single-particle properties}
\label{sec:Results_Electronic}

Figure~\ref{fig:Spectrum_WW_SP} shows the ground-state emission
spectrum without Coulomb interaction for ten different random
configurations. The intensities are all normalized to the maximum
intensity of configuration 5 (Config5). Several interesting features
are clearly visible in the spectra.

Firstly, when looking at Fig.~\ref{fig:Spectrum_WW_SP}, we observe
that different microscopic configurations give significantly
different transition energies. This can also be seen from
Table~\ref{tab:Tranistion_energies}, where the single-particle
transition energies $E^{0}_\text{GS}$ are summarized. Without
Coulomb effects, the difference between the lowest (Config3) and the
highest (Config1) transition energy is 128.7 meV.

Secondly, a striking observation is that six out of ten
configurations have an oscillator strength of less than 20\% of the
oscillator strength of configuration 5 (Config5). This indicates a
weaker spatial wave function overlap of ground-state electron and
hole levels. The main contribution to the emission spectrum arises
here from the configurations 1, 4, 5, and 7. But even these four
configurations have a spread of 73.7 meV in their transition
energies.

\begin{table}[b]
\caption{Ground-state transition energies with ($E_\text{GS}^{X}$)
and without ($E_\text{GS}^{0}$) Coulomb effects included for the
different configurations. For each configuration, the excitonic
binding energy $E_X$ is given.}
\begin{ruledtabular}
\begin{tabular}{|c|ccc|}
\hline
\textbf{Config.} & $E_\text{GS}^{0}$ (eV) & $E_\text{GS}^{X}$ (eV) & $E_X$ (meV)\\
1 & 2.0344 & 2.0017 & 32.7 \\
2 & 2.0119 & 1.9835 & 28.4 \\
3 & 1.9057 & 1.8789 & 26.8 \\
4 & 2.0184 & 1.9850 & 33.4 \\
5 & 1.9607 & 1.9289 & 31.8 \\
6 & 1.9198 & 1.8905 & 29.3 \\
7 & 1.9997 & 1.9642 & 35.5 \\
8 & 1.9560 & 1.9295 & 26.5 \\
9 & 2.0149 & 1.9881 & 26.8 \\
10 & 1.9431 & 1.9182 & 24.9 \\\hline
Average & 1.9765 & 1.9469 & 29.6\\
\end{tabular}
\end{ruledtabular}
\label{tab:Tranistion_energies}
\end{table}

\begin{figure*}[t]
\includegraphics{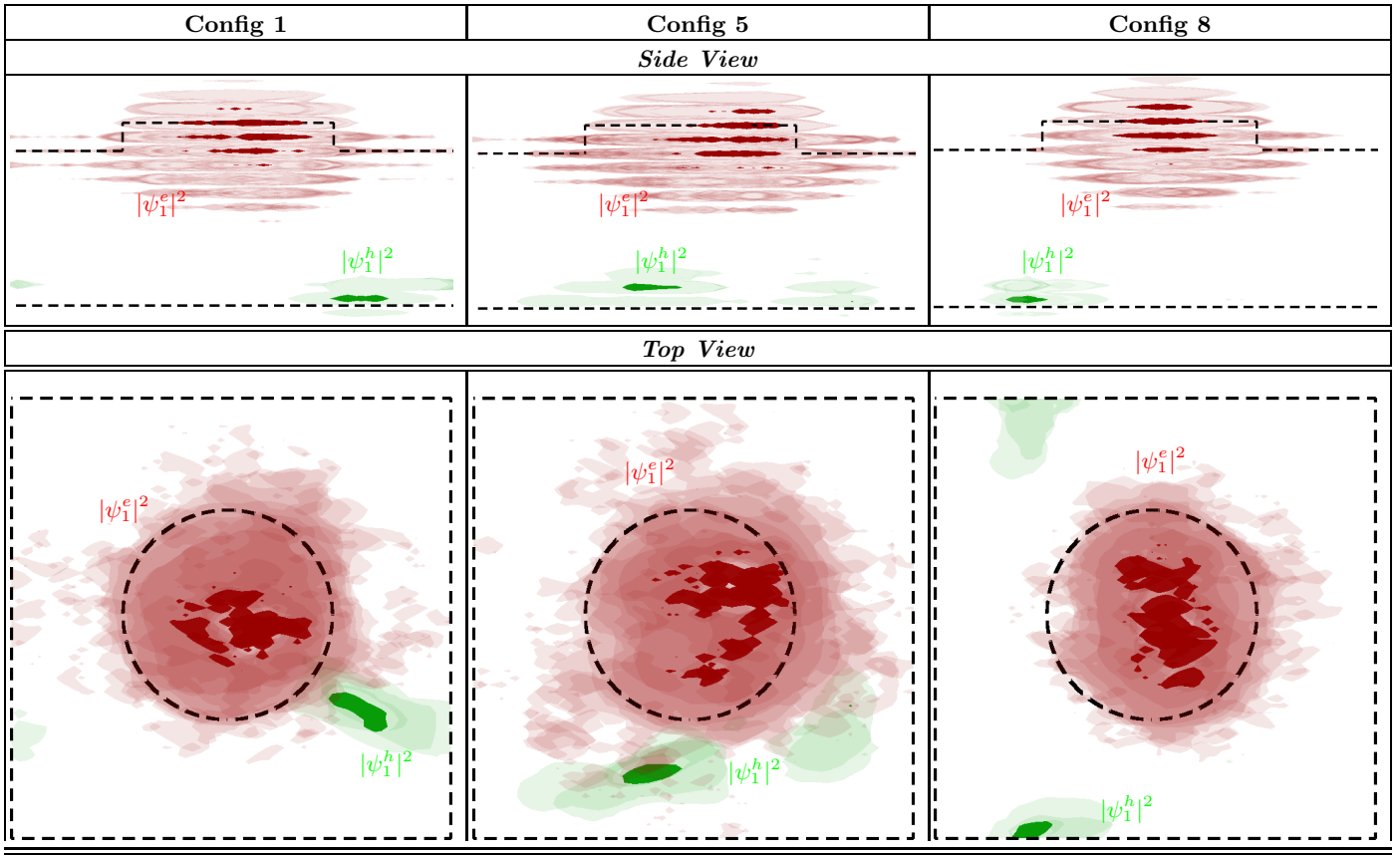}
\caption{(Color online) Ground-state electron (red) and hole (green)
charge densities \emph{without} Coulomb interaction for different
configurations and different view points [first row (side view)
$\perp c$ axis; second row (top view): $\parallel c$ axis]. Light
(dark) isosurfaces correspond to 5\% (50\%) of the maximum charge
density value. Dashed lines indicate the QW interfaces.}
\label{fig:WaveF_SP}
\end{figure*}

The broadening of the PL peak is usually attributed to wave function
localization effects. To gain insight into the microscopic origin of
such wave function localization effects and to elucidate the
difference between different microscopic configurations,
Fig.~\ref{fig:WaveF_SP} shows the calculated single-particle
electron and hole ground-state charge densities $|\psi^{e}_1|^{2}$
and $|\psi^{h}_1|^{2}$, respectively,  for Config1, Config5, and
Config8. The charge densities for electrons (holes) are shown in red
(green). A top view ($\parallel c$ axis) and a side view ($\perp c$
axis) are given. The dashed lines indicate approximately the QW
interfaces. The dark isosurfaces correspond to 50 \% of the maximum
value of the charge density, while the light isosurfaces correspond
to 5\% of the maximum value. Based on Fig.~\ref{fig:Spectrum_WW_SP},
these configurations have been chosen to represent the situation of
a configuration with (i) a very large oscillator strength (Config
5), (ii) a very low oscillator strength (Config 8), and (iii) a
system (Config 1) that can be regarded as an intermediate situation
between (i) and (ii).

These charge density plots have several features in common. Firstly,
the strong electrostatic built-in field leads to a spatial
separation of electron and hole wave functions along the $c$ axis.
However, and in contrast to standard continuum-based descriptions,
which treat InGaN/GaN QWs as homogenous structures described by
average parameters, we find a very strong (nm scale) hole wave
function localization. Our results indicate therefore that already a
random alloy is sufficient to lead to strong hole wave function
localization effects. The electron wave functions are mainly
localized by the presence of the WWF and show a larger localization
length. It should also be noted that the electron wave functions are
affected by local effects, since the charge density does not display
a circular symmetry within the WWF. The localization length is here,
to a first approximation, given by the dimensions of the WWF. The
difference in the observed localization behavior can be attributed
to the much higher hole effective mass compared to the electron
effective mass.~\cite{RiWi2008} Thus a much stronger hole wave
function localization could be expected, consistent with our
calculations. Our findings are in agreement with the results
reported in Ref.~\onlinecite{WaGo2011} for the hole states, but show
that the electron charge density is also impacted by alloy
fluctuation effects. Additionally, both electron and hole charge
densities reflect the anion-cation structure of the underlying WZ
lattice, with hole (electron) states preferentially located at anion
(cation) planes/sites.

There are also differences clearly visible between the chosen
configurations. In contrast to configurations 1 and 5, the hole wave
function in configuration 8 is localized far away from the well
width fluctuation (localization region of the electron) in
\emph{and} perpendicular to the $c$ plane. Consequently, one is left
with a very weak spatial overlap and thus a very low oscillator
strength. Configurations 1 and 5, however, are very similar with the
hole wave function localizing near the WWF in the $c$ plane.
However, configuration 5 seems to have a slightly higher overlap in
the $c$ plane in comparison with configuration 1. This observation
is also reflected in the transition strength displayed in
Fig.~\ref{fig:Spectrum_WW_SP}.

\begin{figure}[t]
\includegraphics{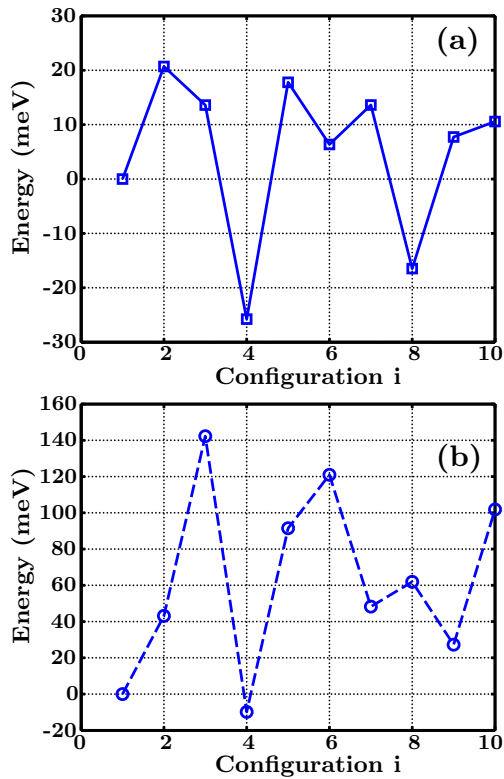}
\caption{Variation of the electron and hole ground-state energies as
a function of the microscopic configuration $i$. All energies are
given with respect to the ground-state energies of configuration 1
($\Delta
E_\text{GS}^{e,h}=E_{\text{GS},i}^{e,h}-E_{\text{GS},1}^{e,h})$.
Results for the electron ground state energies are shown in (a),
while (b) shows the results for the hole ground state energies.}
\label{fig:GS_Vary}
\end{figure}

\begin{table}
\caption{Orbital character of the hole ground state wave function.
The results are given for the different configurations. The
dominating contributions are indicated in bold.}
\begin{ruledtabular}
\begin{tabular}{|c|cccc||cccc|}
 & \multicolumn{8}{c|}{\textbf{Orbital Contribution (\%)}}\\\hline
 & \multicolumn{4}{c||}{\textbf{Hole}} & \multicolumn{4}{c|}{\textbf{Electron}}\\
  & \multicolumn{4}{c||}{\textbf{Ground State}} & \multicolumn{4}{c|}{\textbf{Ground State}}\\\hline
\textbf{Config.} & $p_x$ & $p_y$ & $p_z$ & $s$ & $p_x$ & $p_y$ & $p_z$ & $s$ \\
1 & \textbf{82.29} & 15.73 & 1.63 & 0.35 & 1.15 & 1.10 & 4.57 & \textbf{93.18} \\
2 & \textbf{75.67} & 20.85 & 2.82 & 0.66 & 0.67 & 0.85 & 4.46 & \textbf{94.02} \\
3 & 27.19 & \textbf{69.48} & 2.69 & 0.64 & 0.88 & 1.07 & 4.79 & \textbf{93.26} \\
4 & \textbf{85.90} & 10.98 & 2.50 & 0.62 & 1.28 & 1.35 & 4.66 & \textbf{92.71} \\
5 & \textbf{73.55} & 24.49 & 1.60 & 0.36 & 1.04 & 1.03 & 4.63 & \textbf{93.29} \\
6 & 17.33 & \textbf{79.36} & 2.66 & 0.65 & 1.05 & 1.11 & 4.68 & \textbf{93.16} \\
7 & \textbf{87.83} & 9.35 & 2.26 & 0.56 & 0.88 & 0.91 & 4.61 & \textbf{93.60} \\
8 & 41.63 & \textbf{55.19} & 2.55 & 0.63 & 1.54 & 1.19 & 4.75 & \textbf{92.52} \\
9 & 14.71 & \textbf{83.02} & 1.76 & 0.51 & 1.09 & 0.89 & 4.66 & \textbf{93.36} \\
10 & 22.66 & \textbf{74.82} & 2.00 & 0.51 & 1.17 & 0.95 & 4.74 & \textbf{93.14} \\
\hline Average & \textbf{52.88} & 44.33 & 2.25 & 0.55 & 1.08 & 1.05
& 4.66 & \textbf{93.22}
\end{tabular}
\end{ruledtabular}
\label{tab:OrbitalCharacter}
\end{table}

Having discussed that electron and hole wave functions are strongly
localized for different reasons, we focus in the next step on the
variation of the electron and hole ground-state energies between
different configurations. This analysis reveals which of the two
carrier types determines the variation in the transition energies
shown in Fig.~\ref{fig:Spectrum_WW_SP}. The variation of the
electron and hole ground state energies is displayed in
Fig.~\ref{fig:GS_Vary} (a) and (b), respectively. Here the energies
are calculated with respect to the ground-state energy of
configuration 1 ($\Delta E^{e,h}_\text{GS}(\text{Config}
i)=E^{e,h}_\text{GS}(\text{Config}
i)-E^{e,h}_\text{GS}(\text{Config} 1)$). When looking at
Fig.~\ref{fig:GS_Vary}, in terms of its energy, configuration 1
seems to be an average configuration for the electrons. For the
holes, it is a more extreme case, since the energy difference is
most of the times very large. More specifically, we find that the
electron energies vary in the range of 2-45 meV, while the hole
ground state energies for different microscopic configurations
scatter between 10 and 150 meV. For $\Delta E^{e}_\text{GS}$, the
standard deviation is $\sigma^{e}_\text{GS}=14.23$ meV. For the
holes, we find $\sigma^{h}_\text{GS}=48.08$ meV. This reflects the
much stronger dependence of the hole ground-state energies on the
specific microscopic random alloy configuration. However, it should
be noted that we have considered here a specific size and shape for
the WWF. To shed more light on the impact of WWFs on $\Delta
E^{e,h}_\text{GS}$, we have performed calculations without WWFs.
Since the hole states are localized by the random alloy
fluctuations, when removing the WWF, the spread in $\Delta
E^{h}_\text{GS}$ is similar to the situation with WWFs. However, the
interband transition energies are modified when we exclude WWFs.
This arises from the fact that the electrons are now effectively
localized in a narrower well, with a reduced potential drop across
the well, as illustrated in Fig.~\ref{fig:potential}, where the
total potential drop is calculated to decrease by about 40 meV when
the WWFs are removed. The distribution of WWF sizes and heights in
actual samples should therefore also contribute to the
experimentally observed broadening of InGaN QW absorption and
emission spectra.  A full treatment of the influence of WWFs on the
transition energy would require a range of calculations, taking
account both of variations in WWF sizes and also of the general
reduction in electron-hole overlap with increasing well width and
potential drop. This is beyond the scope of the present study. We
can, however, conclude that random alloy fluctuations predominantly
affect the hole ground-state energies, with WWFs affecting the
electrons states, and with both factors then contributing to the
overall broadening of the experimental absorption and emission
spectra.

\begin{figure}[t]
\includegraphics{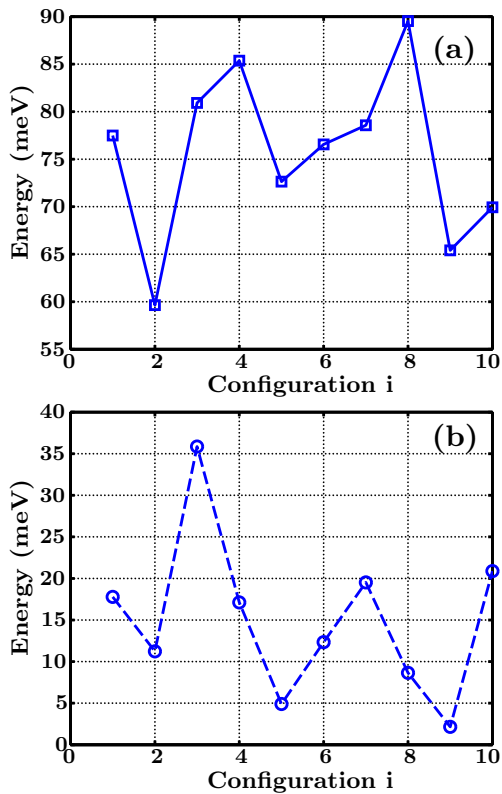}
\caption{(a) Splitting between the electron ground state and first
excited state as a function of the microscopic configuration $i$.
(b) Same as in (a) but here for the hole states.}
\label{fig:splitting}
\end{figure}

In addition, since our analysis reveals that local alloy effects
significantly modify the valence band structure, we study in a
further step the orbital character of electron and hole ground
states. This analysis is important when investigating the optical
polarization characteristic of InGaN/GaN QWs. This study goes beyond
the capabilities of a single-band effective mass approximation. We
note that with standard continuum-based approaches, which treat
$c$-plane InGaN QWs as homogenous structures described by average
parameters, the topmost valence state would have a \emph{charge
density}, independent of the In content, with 50\% $p_x$- and 50\%
$p_y$-like character. Table~\ref{tab:OrbitalCharacter} shows the
contributions of the $p_x$, $p_y$, $p_z$, and $s$ orbitals to the
hole and electron ground state on average and for the different
configurations. The dominant orbital contribution in each
configuration is highlighted. The electron ground state shows the
expected behavior that the wave function is mainly given by
$s$-orbital contributions with much weaker $p_z$ contributions and
negligible $p_x$ and $p_y$ character. For the hole ground state we
observe strong variations in the $p_x$- and $p_y$-like orbital
contributions. The fluctuations in these quantities can be
attributed to local strain effects, explicitly taken into account in
our model. However, \emph{on average} our result is similar to the
result expected from a multiband continuumlike description.

So far, we have focused our attention on ground-state properties.
Since excited states make also a significant contribution to the
optical properties, e.g. when describing the ``S-shape'' of the
temperature dependence of the PL peak
energies,~\cite{WaJi2012,HaWa2012} we turn now to discuss excited
states.

\subsubsection{Excited states}

\begin{figure}[t]
\includegraphics{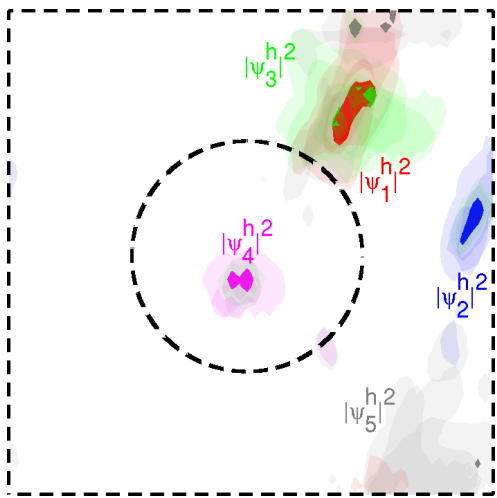}
\caption{(Color online) Single-particle charge densities for the
hole ground state and the first four excited states. The light
(dark) isosurfaces correspond to 5\% (50\%) of the maximum
probability densities. The results are shown for configuration 1.
The dashed lines indicate the QW interfaces.} \label{fig:excited}
\end{figure}

In a first step we look at the energetic separation $\Delta
E^{e}_\text{GS-FE}\text{(Config i)}=E^{e}_\text{FE}\text{(Config
i)}-E^{e}_\text{GS}\text{(Config i)}$ and $\Delta
E^{h}_\text{GS-FE}\text{(Config i)}=E^{h}_\text{GS}\text{(Config
i)}-E^{h}_\text{FE}\text{(Config i)}$, respectively, of the ground
state (GS) and the first excited (FE) state for electrons
$E^{e}_\text{GS-FS}\text{(Config i)}$ and also for holes
$E^{h}_\text{GS-FE}\text{(Config i)}$ as a function of the
configuration $i$. Figure~\ref{fig:splitting} (a) shows $\Delta
E^{e}_\text{GS-FE}\text{(Config i)}$, while Fig.~\ref{fig:splitting}
(b) depicts $\Delta E^{h}_\text{GS-FE}\text{(Config i)}$. From this
figure one can infer again a strong difference between the results
for the electrons and the holes. For the electrons $\Delta
E^{e}_\text{GS-FE}\text{(Config i)}$ scatters between 60-90 meV,
while $\Delta E^{h}_\text{GS-FE}\text{(Config i)}$ varies from 3-35
meV. The calculated mean value for electrons is $\overline{\Delta
E^{e}_\text{GS-FE}}=75.60$ meV and for holes $\overline{\Delta
E^{h}_\text{GS-FE}}=15.04$ meV. The calculated standard deviations
$\sigma^{e}_\text{GS-FE}=8.55$ meV and
$\sigma^{h}_\text{GS-FE}=9.12$ meV are very similar. However, when
looking at $\sigma/\overline{\Delta E_\text{GS-FE}}$, we find a much
smaller value for electrons ($\sigma^{e}/\overline{\Delta
E^{e}_\text{GS-FE}}=0.11$) than for holes
($\sigma^{h}/\overline{\Delta E^{h}_\text{GS-FE}}=0.61$). This
difference can be related to the difference in the impact of local
alloy fluctuations and QW confinement effects on the electronic
structure. We have already seen that the hole wave functions are
significantly affected by local alloy fluctuations, while the
electron wave functions, to a first approximation are localized by
the WWFs. As discussed above, this difference arises from the
difference in the effective masses. Therefore, the electrons are
mainly affected by the overall confinement potential of the QW.
Given the low electron effective mass, compared to the holes, a
large splitting between ground and first excited electron state can
be expected. The hole wave functions, due to their high effective
mass, however, can localize in different potential minima/maxima
originating from alloy fluctuations. This means that not only the
hole ground state is expected to be highly localized but also
excited states. To confirm this we have calculated the charge
densities of the first five hole states for a given random
supercell. As an example we have chosen configuration 1 (Config1)
here, and the results are depicted in Fig.~\ref{fig:excited}. One
can clearly see that not only the hole ground state ($\psi^{h}_1$)
reveals a strong localization but also the shown excited states.
This behavior is \emph{not} a particularity of configuration 1, all
the other configurations show a similar behavior. The presence of
localized excited states is consistent with the experimental
observation of the ``S-shape'' temperature dependence in the PL
spectra of InGaN/GaN QWs.

All the above highlighted factors shed further light on the features
observed in the calculated emission spectrum [cf.
Fig.~\ref{fig:Spectrum_WW_SP}]. The presence of the strong built-in
field combined with the strong hole wave function localization,
arising from local alloy fluctuations, and the electron wave
function localization due to WWFs leads to small wave function
overlaps of electron and hole ground states. Thus, the ground-state
electrons and holes are likely to be localized at spatially
separated positions in \emph{and} perpendicular to the $c$ plane.
This explains that the ground state transition strength in six out
of ten configurations is weak. It should be noted that therefore a
QD-like description of random alloy fluctuations in a
continuum-based model might fail, since it does not account for the
effect that the charge carriers are likely to be localized in
different in-plane spatial positions.

So far, we have not taken into account Coulomb effects, which could
increase the spatial overlap of electron and hole wave functions by
compensating localization and built-in potential effects. Thus, we
focus on the impact of Coulomb effects on the results in the next
section.

\subsection{Coulomb effects}
\label{sec:Results_Optical}

To include Coulomb effects in the description, we use the CI scheme
described in Sec.~\ref{sec:ManyBdy}. Since we have seen in the
previous section that the energetic separation between different
hole states is small, we include the first 15 hole states in the CI
expansion. For the electrons we include the five energetically
lowest single-particle states.

\begin{figure}[t!]
\includegraphics{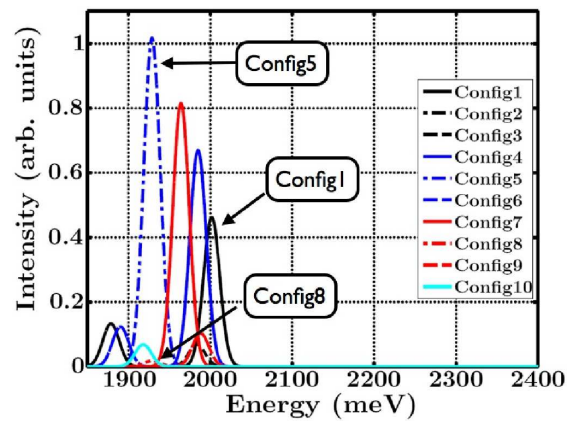}
\caption{(Color online) Excitonic ground-state emission spectrum of
an $\text{In}_{0.25}\text{Ga}_{0.75}$N/GaN QW for different random
configurations including alloy and WWFs.} \label{fig:Spectrum_WW_MB}
\end{figure}

Figure~\ref{fig:Spectrum_WW_MB} shows the calculated excitonic
ground-state emission spectrum for the here considered ten different
random configurations. The intensities are all normalized to the
maximum intensity of configuration 5 \emph{without} Coulomb effects
[cf. Fig.~\ref{fig:Spectrum_WW_SP}]. Compared to
Fig.~\ref{fig:Spectrum_WW_SP}, the (attractive) Coulomb interaction
seems to introduce mainly an energetic shift of the whole emission
spectrum. When including Coulomb effects the oscillator strength is,
for the different transitions, only slightly increased. This
indicates that the spatial separation of the electron and hole wave
functions due to the presence of built-in potential and localization
effects is much stronger than the attractive Coulomb interaction
between the carriers. Similar to the situation without Coulomb
effects we observe that different microscopic configurations give
significantly different transition energies. The excitonic
transition energies $E^{X}_\text{GS}$ are summarized in
Table~\ref{tab:Tranistion_energies}. With Coulomb effects the
difference between the lowest (Config3) and the highest (Config1)
transition energy is 122.8 meV, which is only slightly different to
the result in the absence of the Coulomb interaction (128.7 meV).

\begin{figure*}[t]
\includegraphics{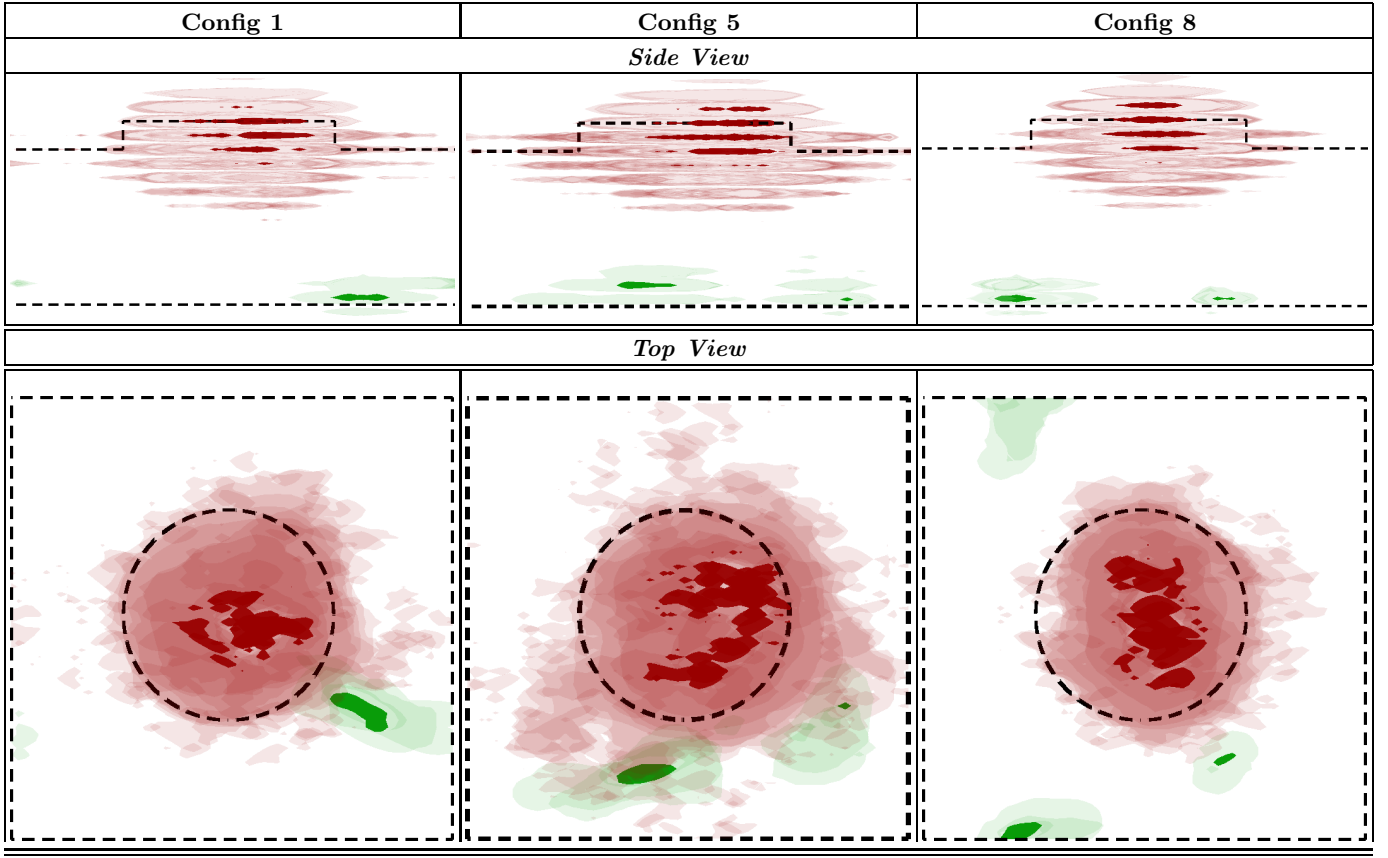}
\caption{(Color online) Ground state electron (green) and hole
(blue) charge densities with Coulomb interaction for different
configurations and different view points [first row (side view)
$\perp c$ axis; second row (top view): $\parallel c$ axis]. Light
(dark) isosurfaces correspond to 5\% (50\%) of the max. charge
density value. Dashed lines indicate the QW interfaces.}
\label{fig:WaveF_MB}
\end{figure*}

Also, Table~\ref{tab:Tranistion_energies} summarizes the calculated
excitonic shifts for the different configurations. On average we
find a shift of 29.6 meV, corresponding to the exciton binding
energy. Recently, Wei and co-workers~\cite{WeJi2012} analyzed the
excitonic binding energy in the framework of a single-band effective
mass description, but neglecting alloy or well-width fluctuations.
The authors find an excitonic binding energy of approximately 25 meV
for a 3.5 nm wide $\text{In}_{0.25}\text{Ga}_{0.75}$N/GaN QW. The
continuum based calculations by Funato and Kawakami,~\cite{FuKa2008}
give for an $\text{In}_{0.25}\text{Ga}_{0.75}$N/GaN QW with 3 nm
width approximately 23 meV. However, these continuum-based QW
calculations neglect the effect of \emph{in-plane} carrier
localization due to alloy fluctuations or WWFs. Such an in-plane
confinement would increase the in-plane wave function overlap and
consequently leads to an increase of the excitonic binding energy.
As discussed before, the single-particle wave functions, as shown in
Fig.~\ref{fig:Spectrum_WW_SP}, reveal not only a strong confinement
along the $c$-axis but also in-plane. While for configurations 1 and
5 electron and hole wave functions are almost localized at the same
in-plane position, in configuration 8 they are not. This is also
reflected in the excitonic binding energies. Configurations 1 and 5
have almost identical excitonic binding energies (32.7 meV versus
31.8 meV), while for configuration 8 the excitonic binding energy is
much smaller (26.5 meV).

Funato and Kawakami~\cite{FuKa2008} calculated also the excitonic
binding energies of cylindrical InGaN/GaN \emph{QDs}, therefore
including lateral confinement effects. In the case of a cylindrical
InGaN/GaN QD with a height of 3 nm and a diameter of 6 nm, the
excitonic binding energies were increased, compared to a QW with an
identical height and composition, by approximately 13 meV. For 25 \%
In, this gives approximately an excitonic binding energy of 36 meV.
This value is comparable to our average excitonic binding energy of
29.6 meV, which also includes, due to alloy and WWFs, lateral
confinement effects.

Having discussed the impact of Coulomb effects on the emission
spectrum, we focus in the next step on the impact of the Coulomb
interaction on the charge densities. In the CI scheme the excitonic
many-body wave function $|\psi^{X}\rangle$ is not a simple product
of electron and hole wave functions, but it is written as a
\emph{linear combination} of electron-hole basis states:
\begin{equation}
|\psi^{X}\rangle=\sum\limits_{i,j}c^{X}_{ij}\hat{e}^{\dagger}_{i}\hat{h}^{\dagger}_j|0\rangle\,
.
\end{equation}
Here, $|0\rangle$ is the vacuum state, $c^{X}_{ij}$ the expansion
coefficient and $\hat{e}^{\dagger}_{i}$ ($\hat{h}^{\dagger}_{i}$)
denotes the electron (hole) creation operator.~\cite{BaSc2013} To
visualize the electron and hole contribution to $|\psi^{X}\rangle$
separately we use a reduced density matrix for electrons and
holes.~\cite{BaSc2013}
For the electrons, the density operator $\hat{\rho}^{e}$ is given
by:
\begin{equation}
\hat{\rho}^{e}=\sum_{i,i'}|i\rangle\sum_{j}c^{X}_{ij}c^{X*}_{i'j}\langle
i'|=\sum_{i,i'}|i\rangle\rho^{e}_{ii'}\langle i'| \,\, .
\end{equation}
One can now calculate the electron and hole densities $\langle
\mathbf{R}|\hat{\rho}^{e}|\mathbf{R}\rangle$ and $\langle
\mathbf{R}|\hat{\rho}^{h}|\mathbf{R}\rangle$, respectively.
Following Fig.~\ref{fig:WaveF_SP}, Fig.~\ref{fig:WaveF_MB} depicts
the calculated electron [$\langle
\mathbf{R}|\hat{\rho}^{e}|\mathbf{R}\rangle$] and hole [$\langle
R|\hat{\rho}^{h}|R\rangle$] densities for the configurations 1, 5,
and 8. When comparing Figs.~\ref{fig:WaveF_SP}
and~\ref{fig:WaveF_MB}, we infer that the position of the charge
densities along the $c$ axis are only slightly affected by the
Coulomb interaction for the here considered
$\text{In}_{0.25}\text{Ga}_{0.75}$N/GaN QW. This shows that the
spatial separation of the charge carriers is dominated by the
built-in field, explaining that the oscillator strength is only very
slightly modified by the attractive Coulomb interaction [cf.
Fig.~\ref{fig:Spectrum_WW_SP}]. This situation could change with
decreasing In content due to reduced built-in fields.

Similar to the effect along the $c$ axis, we find that electron and
hole charge densities are only slightly modified by the Coulomb
interaction in the $c$ plane. Based on the above effective mass
argument, one could expect that the hole state is less affected by
the Coulomb interaction, and that the electron wave function mainly
changes and localizes near the hole. All the displayed
configurations show that localization effects due to random alloy
and well-width fluctuations are stronger than Coulomb effects. Only
slight effects in the shape of the electron and hole wave functions
are visible.

\section{Comparison with experimental data}
\label{sec:Results_Experiment}

Our finding of localization effects on the nanometer scale is in
good agreement with the experimental analysis by Graham \emph{et
al.}~\cite{GrSo2005} Graham and co-workers used the Huang-Rhys
factor to analyze the localization length for the carriers in
InGaN/GaN QWs. For varying In content, the authors estimated
localization lengths of 1.1-3.1 nm.~\cite{GrSo2005} Our results are
consistent with these experimental findings. However, Graham
\emph{et al.} assumed that the localization length is the same for
electrons and holes. Our results reveal that the experimentally
estimated localization length is consistent with the hole wave
function localization, since the localization length of the
electrons will be mainly determined by the size of in-plane WWFs, at
least for the here considered 25\% InN system. It should also be
noted that we have studied here the lower limit (5 nm) of the
experimentally reported values (5-10 nm).~\cite{GaOl2008} Thus, one
could expect that when increasing the in-plane dimension of the
WWFs, the localization length of the electrons is of the order 5-10
nm.

Next, we compare our calculated excitonic transition energies with
available experimental data and analyze also the emission spectrum
in terms of the experimentally available data on the full width half
maximum (FWHM) of the PL peak. Here we have only a small number of
different microscopic random alloy configurations. Thus, a detailed
comparison is difficult. However, the present analysis allows us to
study if our approach gives numbers of the right order of magnitude.
For the here studied 3.5 nm wide
$\text{In}_{0.25}\text{Ga}_{0.75}$N/GaN QW, we estimate an average
excitonic transition energy of $\approx$ 1.95 eV [cf.
Table~\ref{tab:Tranistion_energies}]. Graham \emph{et
al.}~\cite{GrSo2005} extracted for a 3.3 nm wide
$\text{In}_{0.25}\text{Ga}_{0.75}$N/GaN QW a PL peak energy of 2.162
eV, which is in good agreement with our calculations, given that our
well is slightly wider and that we neglect also electron-phonon
coupling effects.

We have also seen in Sec.~\ref{sec:Results_Optical}, that the main
contribution to the calculated excitonic emission spectrum (cf.
Fig.~\ref{fig:Spectrum_WW_MB}) arise from the configurations 1, 4,
5, and 7. The ground-state emission energy for those configurations
is spread over a range of 72.8 meV. This value can now be compared
to the experimental study of the FWHM of the PL peak in
Ref.~\onlinecite{GrSo2005} (cf. Ref.~\onlinecite{WaGo2011}). The
experimental data for a 3.3 nm wide
$\text{In}_{25}\text{Ga}_{0.75}$N/GaN QW gives a FWHM of about 63
meV.

As discussed above, for a more detailed comparison with
experimentally available data, the analysis of more microscopic
structures, different well width, WWFs and In contents would be
required. This is beyond the scope of the present study, since we
are here interested in the discussion of the general features of
$c$-plane InGaN/GaN QWs when taking alloy and WWFs explicitly into
account. Nevertheless, our first results for the localization
length, excitonic transition energies and their energetic variation
give already values in very reasonable agreement with those reported
in literature experimental studies.

\section{Conclusion}
\label{sec:Summary}

In conclusion, we have presented an atomistic approach, including
local alloy, well-width, strain and built-in field fluctuations, to
analyze the electronic and optical properties of $c$-plane
$\text{In}_{0.25}\text{Ga}_{0.75}$N/GaN QWs. Our calculations show
that \emph{random} alloy fluctuations lead to a very strong hole
wave function localization (on the nm scale) which affects the QW
optical properties significantly. The observed hole wave function
localization is consistent with the experimentally estimated
localization lengths of 1-3 nm.~\cite{GrSo2005} Additionally, we
find that the electron wave functions are mainly localized by WWFs,
with some contributions from the local alloy fluctuations. Moreover,
by treating the optical properties in the CI frame, we were able to
show that the emission spectra is dominated by the localized
single-particle states due to well-width and alloy fluctuations and
that the Coulomb interaction mainly affects the carrier wave
functions in the growth plane and leads to overall energetic shifts
of the emission spectrum. However, it should be noted that this
behavior could be significantly different for systems with low In
content or structures grown on non- or semi-polar planes. These
systems will be analyzed in future studies.

\begin{acknowledgments}
The authors would like to thank Philip Dawson, Rachel A. Oliver,
Menno Kappers and Colin J. Humphreys for fruitful discussions. This
work was supported by Science Foundation Ireland (project numbers
10/IN.1/I2994 and 13/SIRG/2210) and the European Union 7th Framework
Programme DEEPEN (grant agreement no.: 604416).
\end{acknowledgments}

\bibliographystyle{apsrev}
\bibliography{../../../phdstef}

\end{document}